\documentclass[a4paper,12pt]{article}

\pdfoutput=1

\usepackage[tbtags]{amsmath}
\usepackage{amssymb}
\usepackage{ifthen}
\usepackage{slashed}
\usepackage{amsfonts}
\usepackage{mathrsfs}
\usepackage{bm}
\usepackage{graphicx,subfigure,booktabs}
\usepackage[numbers,sort&compress]{natbib}
\usepackage{verbatim}
\usepackage{appendix}

\usepackage[colorlinks,
            linkcolor=black,
            filecolor=black,
            anchorcolor=black,
            urlcolor=black,
            citecolor=blue
            ]{hyperref}

\usepackage{color}
\usepackage{ulem}
\usepackage{setspace}
\numberwithin{equation}{section}

\newlength{\dinwidth}
\newlength{\dinmargin}
\makeatletter
\newcommand{\thickhline}{%
    \noalign {\ifnum 0=`}\fi \hrule height 1pt
    \futurelet \reserved@a \@xhline
}
\makeatother

\allowdisplaybreaks

\begin{document}

\title{\bf Revisiting the $P$-wave charmonium radiative decays \boldmath{$h_{c}\rightarrow\gamma\eta^{(\prime)}$} with relativistic corrections}

\author{Jun-Kang He$^{a}$\; and
Chao-Jie~Fan$^{a,b}$\footnote{fancj@hbnu.edu.cn}\\[15pt]
{$^a$\small College of Physics and Electronic Science, Hubei Normal University, Huangshi 435002, China}\\[0.2cm]
{$^b$\small Institute of Particle Physics and Key Laboratory of Quark and Lepton Physics~(MOE), }\\
{\small Central China Normal University, Wuhan, Hubei 430079, China}}
\date{}


\maketitle
\vspace{0.2cm}

\begin{abstract}
{\noindent} The $P$-wave charmonium decays $h_{c}\rightarrow\gamma\eta^{(\prime)}$ are revisited by taking into account relativistic corrections. The decay amplitudes are derived in the Bethe-Salpeter formalism, in which the involved one-loop integrals are evaluated analytically. Intriguingly, from both the quark-antiquark content and the gluonic content of $\eta^{(\prime)}$, the relativistic corrections make significant contributions to the decay rates of $h_{c}\rightarrow\gamma\eta^{(\prime)}$. By comparison with the leading-order contributions from the quark-antiquark content (one-loop level), the ones from the gluonic content (tree level) are also important, which is compatible with the conclusion obtained without relativistic corrections. Usually, for $\eta$ production processes, the predicted branching ratios are sensitive to the angle of $\eta-\eta^{\prime}$ mixing. As an illustration, using the Feldmann-Kroll-Stech result about the mixing angle $\phi=39.3^{\circ}\pm1.0^{\circ}$ as input, we find that the predicted ratio $R_{h_{c}}=\mathcal{B}(h_{c}\rightarrow\gamma\eta)/\mathcal{B}(h_{c}\rightarrow\gamma\eta^{\prime})$ is much smaller than the experiment measurement. While, with $\phi=33.5^{\circ}\pm0.9^{\circ}$ extracted from the asymptotic limit of the $\gamma^{\ast}\gamma-\eta^{\prime}$ transition form factor, we obtain $R_{h_{c}}=30.3\%$ in consistent with $R_{h_{c}}^{exp}=(30.7\pm11.3\pm8.7)\%$. As a cross-check, the mixing angle $\phi=33.8^{\circ}\pm2.5^{\circ}$ is extracted by employing the ratio $R_{h_{c}}$, and a brief discussion on the difference in the determinations of $\phi$ is given.

\end{abstract}

\newpage


\section{Introduction}
\label{sec:intro}
The hadronic decays of charmonia have played important roles for our understanding of quantum chromodynamics (QCD), especially the interplay of perturbative QCD and nonperturbative QCD~\cite{Novikov:1977dq,Brodsky:1981kj,Chernyak:1983ej,Voloshin:2007dx}, since the first charmonium state $J/\psi$ was observed~\cite{Aubert:1974js,Augustin:1974xw}. One of the interesting topics is the Okubo-Zweig-Iizuka
(OZI~\cite{Okubo:1963fa,Zweig:1981pd,Zweig:1964jf,Iizuka:1966fk})-suppressed radiative decays of charmonia to the light mesons $\eta^{(\prime)}$. On the one hand, these decays are closely related to the issue of $\eta-\eta^{\prime}$ mixing, which could shed light on the $U(1)_{A}$ anomaly~\cite{Adler:1969gk,Bell:1969ts,Weinberg:1975ui,Witten:1978bc,Witten:1979vv,
Veneziano:1979ec,Feldmann:1998vh,Feldmann:1999uf,Escribano:2020jdy} and the $SU(3)_{F}$ breaking~\cite{Kazi:1975tu,Fritzsch:1976qc,Feldmann:1998vh,Feldmann:1999uf,Escribano:2020jdy}. On the other hand, these decays provide a relatively clean environment to study the gluonic content of $\eta^{(\prime)}$, since there is no complication of interactions between the final light hadrons.

In recent years, there are more and more experimental measurements on the radiative decays of charmonia to $\eta^{(\prime)}$, such as $J/\psi\rightarrow\gamma\eta^{(\prime)}$~\cite{Ablikim:2005je,Libby:2008fg,Pedlar:2009aa,Ablikim:2010kp}, $\psi^ {\prime}\rightarrow\gamma\eta^{(\prime)}$~\cite{Pedlar:2009aa,Ablikim:2010dx,Ablikim:2017vhb}, $\psi(3770)\rightarrow\gamma\eta^{(\prime)}$~\cite{Pedlar:2009aa} and $h_{c}\rightarrow\gamma\eta^{(\prime)}$~\cite{Ablikim:2016uoc}. In the theoretical aspect, the $S$-wave charmonium decays $J/\psi\,(\psi^{\prime})\rightarrow\gamma\eta^{(\prime)}$ have been investigated in various approaches, such as QCD sum rules~\cite{Novikov:1979uy}, chiral and large $N_{C}$ approach~\cite{Chao:1989pi,Chao:1990im,Gerard:2004gx}, QCD multipole expansion~\cite{Kuang:1990kd}, effective Lagrangian approach~\cite{Chen:2014yta}, perturbative QCD~\cite{Korner:1982vg,Ma:2002ww,Yang:2004wy,Li:2005ug,Li:2007dq,He:2019mpy} and phenomenological models~\cite{Zhao:2010mm,Gerard:2013gya}, and predictions of the branching ratios are compatible with experimental data. Furthermore, by the ratio $R_{J/\psi}=\mathcal{B}(J/\psi\rightarrow\gamma\eta^{\prime})/\mathcal{B}(J/\psi\rightarrow\gamma\eta)$, the angle of $\eta-\eta^{\prime}$ mixing was obtained $\phi=39.0^{\circ}\pm1.6^{\circ}$~\cite{Feldmann:1998vh} with nonperturbative matrix elements $\langle0|G_{\mu\nu}^a\tilde{G}^{a,\mu\nu}|\eta^{(\prime)}\rangle$ and $\phi=33.9^{\circ}\pm0.6^{\circ}$~\cite{He:2019mpy} with perturbative QCD. It is worth noting that the recent lattice calculation of the ETM collaboration gives the mixing angle $\phi=46^{\circ}\pm1^{\circ}\pm3^{\circ}$~\cite{Michael:2013gka,Urbach:2017rvx}, while the UKQCD collaboration obtains $\phi=34^{\circ}\pm3^{\circ}$~\cite{Gregory:2011sg}. The discrepancies in these determinations of the mixing angle might indicate that our understanding of $\eta-\eta^{\prime}$ mixing scheme~\cite{Fritzsch:1976qc,Akhoury:1987ed,Ball:1995zv,Feldmann:1997vc,Leutwyler:1997yr,
Kaiser:1998ds,Feldmann:1998vh,Feldmann:1998sh,Feldmann:1999uf} is incomplete, and further experimental and theoretical investigations are needed to make sense of the $\eta-\eta^{\prime}$ mixing.

Turning to $P$-wave charmonia decays, the physical picture seems more complex, since the higher Fock-state contributions and the relativistic corrections may become important. It is well known that infra-red (IR) divergences are encountered in the color-singlet state contributions for the inclusive $P$-wave charmonia decays with the zero-binding approximation~\cite{Barbieri:1976fp,Barbieri:1980yp,Barbieri:1981xz,Kwong:1987ak}. Although these IR divergences can be removed by considering the higher Fock-state contributions from the point of view of nonrelativistic QCD (NRQCD)~\cite{Bodwin:1992ye,Bodwin:1994jh}, they may imply that the effects beyond those contained in the derivative of the nonrelativistic wave function at the origin $R^{\prime}(0)$ may play a key role. Generally, it should be noted that similar IR divergences do not appear in exclusive $P$-wave charmonia decays~\cite{Kroll:1997vt,Wong:1998rv,Wong:1999dj,Wong:2000rj}. Nevertheless, as pointed out in Refs.~\cite{Keung:1982jb,Huang:1996bk,Kroll:1997vt,Wong:1998rv,Wong:1999dj,Wong:2000rj,Ma:2002eva}, the higher-order contributions, such as the higher Fock-state contributions~\cite{Kroll:1997vt,Wong:1998rv,Wong:1999dj,Wong:2000rj,Ma:2002eva} and the relativistic corrections~\cite{Keung:1982jb,Huang:1996bk,Ma:2002eva}, are still important to exclusive $P$-wave charmonia decays. For the exclusive $P$-wave charmonium decays $h_{c} \rightarrow \gamma\eta^{(\prime)}$, there are a few studies~\cite{Zhu:2016udl,Wu:2017pep,Fan:2019sap} in the theoretical aspect ever since the branching ratios $\mathcal{B}(h_{c}\rightarrow\gamma\eta^{\prime})$ and $\mathcal{B}(h_{c}\rightarrow\gamma\eta)$ are first measured to be, respectively, $(1.52\pm0.27\pm0.29)\times 10^{-3}$ and $(4.7\pm1.5\pm1.4)\times 10^{-4}$ by the BESIII Collaboration~\cite{Ablikim:2016uoc}. In the nonrelativistic limit~\cite{Zhu:2016udl,Fan:2019sap}, the relativistic corrections related to the internal momentum of the $P$-wave charmonium $h_{c}$ have been neglected in the calculation of the decay rates, and all the nonperturbative effects are absorbed in $R_{h_{c}}^{\prime}(0)$ with the Taylor expansion of the hard-scattering amplitudes up to the linear terms. Then it is found that the calculations are IR safe and the predicted branching ratios $\mathcal{B}(h_{c}\rightarrow\gamma\eta^{(\prime)})$ are much smaller than the experimental measurements. Obviously, this indicates that the relativistic corrections or/and the contributions from the higher Fock-state of $h_{c}$ are highly significant. While, from the point of view of NRQCD, the next-to-leading-order Fock-state contributions are suppressed by a relative factor $v^{2}_{c\bar{c}}\alpha_{s}$ in the decays $h_{c} \rightarrow \gamma\eta^{(\prime)}$~\cite{Zhu:2016udl}. So it means that the relativistic corrections are needed in the exclusive $P$-wave charmonium decays $h_{c} \rightarrow \gamma\eta^{(\prime)}$.

One of the major concerns of this paper is to study the relativistic corrections in the exclusive $P$-wave charmonium decays $h_{c} \rightarrow \gamma\eta^{(\prime)}$ by performing an explicit calculation. To make clear these relativistic corrections, the Bethe-Salpeter (B-S) framework~\cite{Mitra:1990av,Bhatnagar:2009jg,Bhatnagar:2013bha} is used to calculate the wave function of $h_{c}$ and the decay amplitudes of $h_{c}\rightarrow\gamma\eta^{(\prime)}$, where the internal momentum of $h_{c}$ is retained in both the soft bound-state wave function and the hard-scattering amplitude. Here, it is worth noting that there are at least two sources of the relativistic corrections. One is from the kinematical corrections which appear in the annihilation amplitudes, and the other is from the dynamical corrections of bound-state wave function itself. For the final light mesons $\eta^{(\prime)}$, light-cone distribution amplitudes (DAs) are adopted because of the large momentum transfer. And the contributions of the quark-antiquark content and those of the gluonic content of $\eta^{(\prime)}$ are both taken into account in our calculations.

In this paper, with the technique of helicity projector, we evaluate analytically the involved one-loop integrals with the internal momentum of $h_{c}$ kept. For the contributions from the quark-antiquark content of $\eta^{(\prime)}$ in the decays $h_{c}\rightarrow\gamma\eta^{(\prime)}$, the relativistic effects mainly originate from the kinematic part of the annihilation amplitudes, especially when the internal momentum of $h_{c}$ makes the propagator near on-shell. For the contributions of the gluonic content of $\eta^{(\prime)}$ in the decays $h_{c}\rightarrow\gamma\eta^{(\prime)}$, the next-to-leading-order effects related to the internal momentum are not substantially suppressed in the major region of the wave function of $h_{c}$, and therefore the corresponding relativistic corrections are extremely important. Furthermore, we find out that the gluonic contributions and the quark-antiquark contributions are comparable, unlike the situation in the heavy vector quarkonium decays $V\rightarrow \gamma\eta^{(\prime)}$~\cite{Baier:1981pm,Ma:2002ww,He:2019mpy} where the gluonic contributions are strongly suppressed due to the special form of the spin structure of their amplitudes. In addition, it is also unlike the phenomenological fits~\cite{Escribano:2007cd,Escribano:2008rq,Ambrosino:2009sc} where the gluonic content of $\eta$ can be neglected. This signifies that the decays $h_{c}\rightarrow\gamma\eta^{(\prime)}$ can be used to test the gluonic content of the $\eta^{(\prime)}$ more efficiently than the decay processes $V\rightarrow \gamma\eta^{(\prime)}$. It is worthwhile to point out that the decay rates of $h_{c}\rightarrow\gamma\eta^{(\prime)}$ are insensitive to the light quark masses and the shapes of the $\eta^{(\prime)}$ DAs~\cite{Fan:2019sap}, so the theoretical uncertainties from the $\eta^{(\prime)}$ DAs are negligible, and the mixing angle of the $\eta-\eta^{\prime}$ system could be reliably extracted in our calculations.

The paper is organized as follows. The formalism for the decays $h_{c}\rightarrow\gamma\eta^{(\prime)}$ is presented in section~\ref{sec:framework}. In section~\ref{sec:numerical analysis}, we obtain our numerical results, and the final section is our summary. The expressions of the numerators involved in section~\ref{sec:framework} are given in the Appendix.

\section{Formalism for radiative decays $h_{c}\rightarrow\gamma\eta^{(\prime)}$}
\label{sec:framework}
\subsection{Bethe-Salpeter equation}
\label{subsec:BSE}

It is generally known that the B-S equation~\cite{Salpeter:1951sz,Salpeter:1952ib} is an effective relativistic equation describing a bound state and has a solid basis in quantum field theory. So it is a conventional approach to treat various relativistic bound-state problems. In this subsection, we briefly review the formulation of the B-S framework. For charmonia,
the B-S equation has the form~\cite{Mengesha:2011pu,Negash:2015hma,Negash:2015rua,Bhatnagar:2016otj}
\begin{equation}\label{bse}
S^{-1}_{F}(f)\Psi(K,q)S^{-1}_{F}(-\bar{f}) =
\int\frac{\mathrm{d}^{4}q^{\prime}}{(2\pi)^{4}}\Big{[}-i\mathcal{K}(K,q,q^{\prime})\Psi(K,q^{\prime})\Big{]},
\end{equation}
where $\mathcal{K}(K,q,q^{\prime})$ represents the interaction kernel between the internal quark and antiquark, and $S_{F}(p)=i/(\slashed{p}-\hat{m}_{c}+i\epsilon)$ represents the propagator with
the effective mass of $c$ quark $\hat{m}_{c}$. The momenta of the quark and antiquark can be written as
\begin{equation}
f=\frac{K}{2}+q,~~~~\bar{f}=\frac{K}{2}-q,
\end{equation}
where $q$ and $K$ represent the internal momentum and the total momentum of the charmonia respectively.

For convenience, one can divide the internal momentum $q$ into two parts. One part is the transverse component $\hat{q}$ with $\hat{q}\cdot K=0$, and the other is the longitudinal component $q_{\parallel}$ which is parallel to the total momentum $K$:
\begin{eqnarray}
q^{\mu}&=& q_{\parallel}^{\mu}+\hat{q}^{\mu}, \nonumber\\
q_{\parallel}^{\mu} &=& \frac{q_{K}}{M} K^{\mu}.
\end{eqnarray}
Here both $q_{K}=\frac{q\cdot K}{M}$ and $\hat{q}^{2}=q^{2}-q_{K}^{2}$ are Lorentz invariant variables and $M$ is the mass of the charmonia.

In the rest frame of the charmonia and under the covariant instantaneous ansatz (CIA)~\cite{Mitra:1990av,Bhatnagar:2009jg,Bhatnagar:2013bha}, the interaction kernel $\mathcal{K}(K,q,q^{\prime})$ is taken to be dependent only on the momentum $\hat{q}$
\begin{equation}\label{k}
\mathcal{K}(K,q,q^{\prime})=V(\hat{q},\hat{q}^{\prime}).
\end{equation}
Generally, the interquark potential $V(\hat{q},\hat{q}^{\prime})$ include both the long-ranged confinement and the short-ranged one-gluon exchange interactions~\cite{Chao:1995cz,Chao:1996bj,Li:2009zu,Li:2009nr,Bhatnagar:2016otj,Gebrehana:2019mpw}.

Then the B-S wave function can be expressed as
\begin{equation}\label{bsecia}
\Psi(K,q)=-S_{F}(f)\Gamma(\hat{q})S_{F}(-\bar{f}),
\end{equation}
where the hadron-quark vertex function reads
\begin{equation}
   \Gamma(\hat{q})=i\int\frac{\mathrm{d}^{4}q^{\prime}}
   {(2\pi)^{4}}\mathcal{K}(\hat{q},\hat{q}^{\prime})\Psi(K,q^{\prime})=\int\widetilde{\mathrm{d}^{3}q^{\prime}}
   V(\hat{q},\hat{q}^{\prime})\psi(\hat{q}^{\prime})
\end{equation}
with the Salpeter wave function $\psi(\hat{q})=\frac{i}{2\pi}\int\mathrm{d}q_{K}\Psi(K,q)$ and $\widetilde{\mathrm{d}^{3}q^{\prime}}=\frac{\mathrm{d}^{3}q^{\prime}}{(2\pi)^{3}}$. Using the operators~\cite{Negash:2015hma,Negash:2015rua,Bhatnagar:2016otj,Gebrehana:2019mpw}
\begin{equation}
  \Lambda^{\pm}_{i}(\hat{q})=\frac{1}{2\omega}\left[\frac{\slashed{K}}{M}\omega
  \pm(-1)^{(i+1)}(\hat{m}_{c}+\slashed{\hat{q}})\right],
\end{equation}
the propagators can be decomposed as
\begin{eqnarray}
\frac{1}{\slashed{f}-\hat{m}_{c}+i\epsilon }&=&\frac{\Lambda^{+}_{1}(\hat{q})}{q_{K}+\frac{M}{2}-\omega+i\epsilon}
  +\frac{\Lambda^{-}_{1}(\hat{q})}{q_{K}+\frac{M}{2}+\omega-i\epsilon},\nonumber\\
 \frac{1}{\slashed{\bar{f}}+\hat{m}_{c}-i\epsilon}&=& \frac{\Lambda^{+}_{2}(\hat{q})}{-q_{K}+\frac{M}{2}-\omega+i\epsilon}
  +\frac{\Lambda^{-}_{2}(\hat{q})}{-q_{K}+\frac{M}{2}+\omega-i\epsilon}
\end{eqnarray}
with $\omega=\sqrt{\hat{m}_{c}^{2}-\hat{q}^{2}}$. Performing the $q_{K}$-integration of Eq.~\eqref{bsecia},
one can obtain~\cite{Negash:2015hma,Negash:2015rua,Bhatnagar:2016otj,Gebrehana:2019mpw}
\begin{eqnarray}
(M-2\omega)\psi^{++}(\hat{q})&=&-\Lambda^{+}_{1}(\hat{q})\Gamma(\hat{q})\Lambda^{+}_{2}(\hat{q}),\nonumber \\
(M+2\omega)\psi^{--}(\hat{q})&=&\Lambda^{-}_{1}(\hat{q})\Gamma(\hat{q})\Lambda^{-}_{2}(\hat{q}),\nonumber \\
\psi^{+-}(\hat{q})&=&0,\nonumber\\
\psi^{-+}(\hat{q})&=&0
\end{eqnarray}
with $\psi^{\pm\pm}(\hat{q})=\Lambda^{\pm}_{1}(\hat{q})\frac{\slashed{P}}{M}\psi(\hat{q})
\frac{\slashed{P}}{M}\Lambda^{\pm}_{2}(\hat{q})$ and
$\psi(\hat{q})=\psi^{++}(\hat{q})+\psi^{+-}(\hat{q})+\psi^{-+}(\hat{q})+\psi^{--}(\hat{q})$.

For the axial vector meson $h_{c}$,  the Salpeter wave function can be approximately written as~\cite{LlewellynSmith:1969az,Sauli:2011aa,Sauli:2011wh,Negash:2015rua,Bhatnagar:2016otj}
\begin{eqnarray}
\psi(\hat{q})=\hat{q}\cdot\varepsilon(K) \left[1+\frac{\slashed{K}}{M}+\frac{\hat{\slashed{q}}\slashed{K}}
{\hat{m}_{c}M}\right]\gamma^{5}f(\hat{q}^{2}),
\end{eqnarray}
where $M$ and $\varepsilon(K)$ are the mass and the polarization vector of $h_{c}$ respectively, and the front factor $\hat{q}\cdot\varepsilon(K)$ indicates that the wave function is of $P$-wave nature mainly, and  $f(\hat{q}^{2})$ is a scalar function of $\hat{q}^{2}$. In the rest frame of $h_{c}$, the momenta $K$ and $\hat{q}$ have the form
\begin{eqnarray}
K^{\mu}&=&(M,\mathbf{0}),\quad\quad   \hat{q}^{\mu}=(0,\mathbf{\hat{q}})=(0,\mathbf{q}),
\end{eqnarray}
and the scalar function $f(\hat{q}^{2})$ satisfies the harmonic oscillator equation (more details and discussions could be found in Refs.~\cite{Bhatnagar:2016otj,Gebrehana:2019mpw}). The expression of
$f(\hat{q}^{2})$ reads
\begin{eqnarray}\label{fhatq}
f(\hat{q}^{2})&=&N_{A}\left(\frac{2}{3}\right)^{\frac{1}{2}}\frac{1}{\pi^{\frac{3}{4}}\beta^{\frac{5}{2}}_{{A}}}
\left|\mathbf{q}\right|e^{-\frac{\mathbf{q}^{2}}{2\beta^{2}_{A}}},
\end{eqnarray}
where $N_{A}$ is the normalization constant and $\beta_{A}$ is the harmonic oscillator parameter~\cite{Bhatnagar:2016otj,Gebrehana:2019mpw}. The normalization equation of  $f(\hat{q}^{2})$ reads
\begin{eqnarray}
\int\frac{\mathrm{d}^{3}q}{(2\pi)^{3}}\frac{4\omega \mathbf{q}^{2}}{3\hat{m}_{c} M}f^{2}(\hat{q}^{2})=1.
\end{eqnarray}

\subsection{The contributions of the quark-antiquark content of $\eta^{(\prime)}$}
\label{subsec:QCDq}

\begin{figure}[th]
  \begin{center}
  \includegraphics[width=0.45\textwidth]{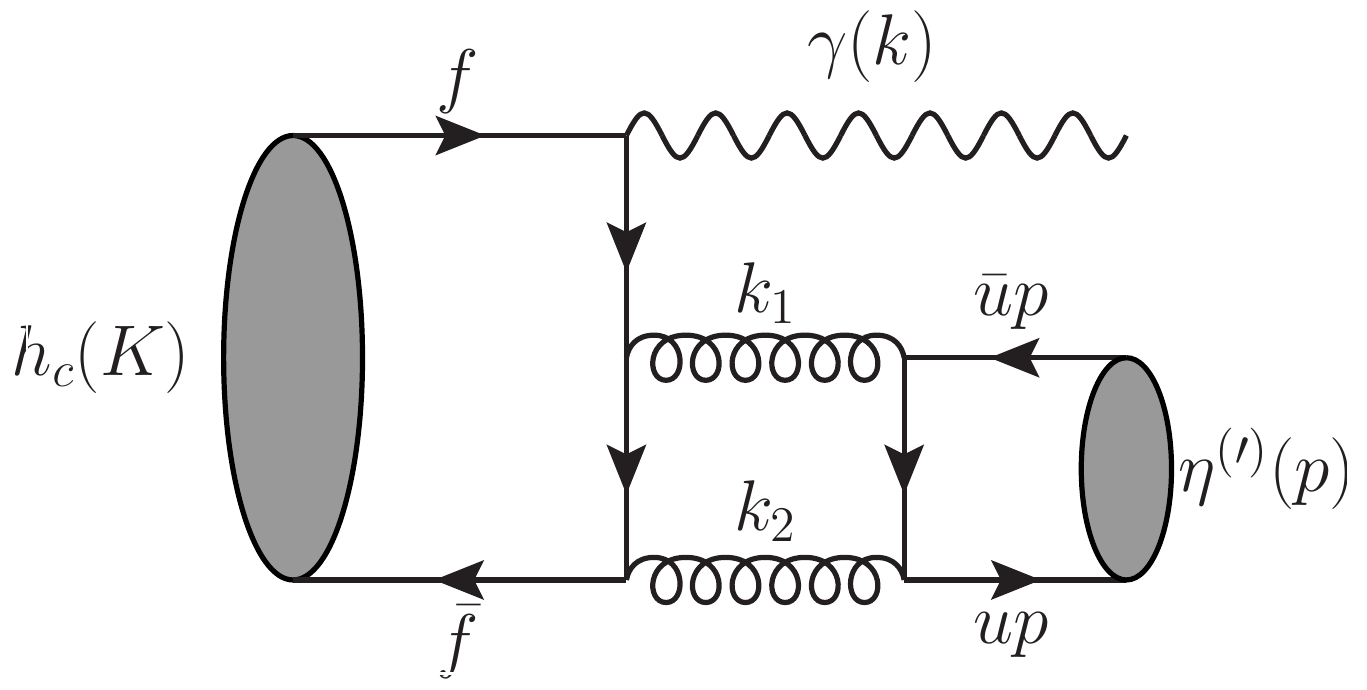}
  \end{center}
  \vskip -0.7cm
  \caption{One typical Feynman diagram for $h_{c}\rightarrow\gamma\eta^{(\prime)}$ with the quark-antiquark content of $\eta^{(\prime)}$. Here the kinematic variables are labeled.}\label{hQCDq}
\end{figure}

For the quark-antiquark content of $\eta^{(\prime)}$, one of the leading-order
Feynman diagrams for the radiative decays $h_{c}\rightarrow\gamma\eta^{(\prime)}$ is depicted in Fig.~\ref{hQCDq}.
Other five diagrams arise from permutations of the photon and gluon legs. And it is convenient to divide the amplitude of $h_{c}\rightarrow\gamma\eta^{(\prime)}$ into two parts~\cite{Fan:2019sap}.
One part describes the effective coupling between $h_{c}$, a real photon and two virtual
gluons, and the other part describes the effective coupling between $\eta^{(\prime)}$ and two
virtual gluons.

In the rest frame of $h_{c}$, the amplitude of $h_{c}\rightarrow\gamma g^{\ast}g^{\ast}$ has the form~\cite{Guberina:1980dc,Guberina:1980xb,Korner:1982vg,Resag:1993xq,Fan:2019sap}
\begin{eqnarray}
{\mathcal A}^{\alpha\beta\mu\nu}\varepsilon_{\alpha}(K)\epsilon^{*}_{\beta}(k)\epsilon^{*}_{\mu}(k_{1})\epsilon^{*}_{\nu}(k_{2})
&=&\sqrt{3}\int\frac{\mathrm{d}^{4}q}{(2\pi)^{4}}\textrm{Tr}\left[\Psi(K,q){\cal O}(f,\bar{f})\right],
\end{eqnarray}
where $k$, $k_{1}$, $k_{2}$ and $\epsilon(k)$, $\epsilon(k_{1})$, $\epsilon(k_{2})$ stand for the momenta and polarization vectors of the photon and the gluons respectively, the factor $\sqrt{3}$ is included to account for the color properties of the quark-antiquark content, ${\cal O}(f,\bar{f})$ is the hard-scattering amplitude, and the momenta of the quark and antiquark read
\begin{eqnarray}
f^{\mu}&=&\frac{K^{\mu}}{2}+q^{\mu}
=\left(\frac{M}{2}+q^{0},\mathbf{q}\right),\nonumber\\
\bar{f}^{\mu}&=&\frac{K^{\mu}}{2}-q^{\mu}
=\left(\frac{M}{2}-q^{0},-\mathbf{q}\right).
\end{eqnarray}
Under the CIA, a more relevant treatment is to take $q^{0}\ll M$, so one can obtain the momenta
\begin{eqnarray}
f^{\mu}&\approx&\left(\frac{M}{2},\mathbf{q}\right)=\frac{K^{\mu}}{2}+\hat{q}^{\mu},\nonumber\\
\bar{f}^{\mu}&\approx&\left(\frac{M}{2},-\mathbf{q}\right)=\frac{K^{\mu}}{2}-\hat{q}^{\mu},
\end{eqnarray}
and the hard-scattering amplitude
\begin{eqnarray}
{\cal O}(f,\bar{f})&\approx&{\cal O}(\hat{q}).
\end{eqnarray}
From another point of view~\cite{Chao:1995cz}, this treatment can be connected with the on-shell condition, which maintains the gauge invariance of the hard-scattering amplitude.

Then the amplitude of $h_{c}\rightarrow\gamma g^{\ast}g^{\ast}$ can be written as
\begin{eqnarray}
{\mathcal A}^{\alpha\beta\mu\nu}\varepsilon_{\alpha}(K)\epsilon^{*}_{\beta}(k)\epsilon^{*}_{\mu}(k_{1})\epsilon^{*}_{\nu}(k_{2})
&=&\sqrt{3}\int\frac{\mathrm{d}^{4}q}{(2\pi)^{4}}\textrm{Tr}\left[\Psi(K,q){\cal O}(\hat{q})\right]\nonumber\\
&=&-i\sqrt{3}\int\frac{\mathrm{d}^{3}\hat{q}}{(2\pi)^{3}}\textrm{Tr}\left[\psi
(\hat{q}){\cal O}(\hat{q})\right],
\end{eqnarray}
where the hard-scattering amplitude ${\cal O}(\hat{q})$ reads
\begin{eqnarray}
{\cal O}(\hat{q})=iQ_{c}eg^{2}_{s}
\frac{\delta_{ab}}{6}&&\Bigg{[}\slashed{\epsilon}^{\ast}(k_{2})\frac{\frac{\slashed{k}_{2}-\slashed{k}-\slashed{k}_{1}}{2}
+\hat{\slashed{q}}+m_{c}}{\left(\frac{k_{2}-k-k_{1}}{2}+\hat{q}\right)^{2}-m^{2}_{c}}
\slashed{\epsilon}^{\ast}(k)\frac{\frac{\slashed{k}_{2}
+\slashed{k}-\slashed{k}_{1}}{2}+\hat{\slashed{q}}+m_{c}}
{\left(\frac{k_{2}+k-k_{1}}{2}+\hat{q}\right)^{2}-m^{2}_{c}}\slashed{\epsilon}^{\ast}(k_{1})\nonumber\\
& &+\slashed{\epsilon}^{\ast}(k_{1})\frac{\frac{\slashed{k}_{1}-\slashed{k}-\slashed{k}_{2}}{2}
+\hat{\slashed{q}}+m_{c}}{\left(\frac{k_{1}-k-k_{2}}{2}+\hat{q}\right)^{2}-m^{2}_{c}}
\slashed{\epsilon}^{\ast}(k)\frac{\frac{\slashed{k}_{1}
+\slashed{k}-\slashed{k}_{2}}{2}+\hat{\slashed{q}}+m_{c}}
{\left(\frac{k_{1}+k-k_{2}}{2}+\hat{q}\right)^{2}-m^{2}_{c}}\slashed{\epsilon}^{\ast}(k_{2})\nonumber\\
& &+\slashed{\epsilon}^{\ast}(k_{2})\frac{\frac{\slashed{k}_{2}-\slashed{k}_{1}-\slashed{k}}{2}
+\hat{\slashed{q}}+m_{c}}{\left(\frac{k_{2}-k_{1}-k}{2}+\hat{q}\right)^{2}-m^{2}_{c}}
\slashed{\epsilon}^{\ast}(k_{1})\frac{\frac{\slashed{k}_{2}
+\slashed{k}_{1}-\slashed{k}}{2}+\hat{\slashed{q}}+m_{c}}
{\left(\frac{k_{2}+k_{1}-k}{2}+\hat{q}\right)^{2}-m^{2}_{c}}\slashed{\epsilon}^{\ast}(k)\nonumber\\
& &+\slashed{\epsilon}^{\ast}(k)\frac{\frac{\slashed{k}-\slashed{k}_{2}-\slashed{k}_{1}}{2}
+\hat{\slashed{q}}+m_{c}}{\left(\frac{k-k_{2}-k_{1}}{2}+\hat{q}\right)^{2}-m^{2}_{c}}
\slashed{\epsilon}^{\ast}(k_{2})\frac{\frac{\slashed{k}
+\slashed{k}_{2}-\slashed{k}_{1}}{2}+\hat{\slashed{q}}+m_{c}}
{\left(\frac{k+k_{2}-k_{1}}{2}+\hat{q}\right)^{2}-m^{2}_{c}}\slashed{\epsilon}^{\ast}(k_{1})\nonumber\\
& &+\slashed{\epsilon}^{\ast}(k_{1})\frac{\frac{\slashed{k}_{1}-\slashed{k}_{2}-\slashed{k}}{2}
+\hat{\slashed{q}}+m_{c}}{\left(\frac{k_{1}-k_{2}-k}{2}+\hat{q}\right)^{2}-m^{2}_{c}}
\slashed{\epsilon}^{\ast}(k_{2})\frac{\frac{\slashed{k}_{1}
+\slashed{k}_{2}-\slashed{k}}{2}+\hat{\slashed{q}}+m_{c}}
{\left(\frac{k_{1}+k_{2}-k}{2}+\hat{q}\right)^{2}-m^{2}_{c}}\slashed{\epsilon}^{\ast}(k)\nonumber\\
& &+\slashed{\epsilon}^{\ast}(k)\frac{\frac{\slashed{k}-\slashed{k}_{1}-\slashed{k}_{2}}{2}
+\hat{\slashed{q}}+m_{c}}{\left(\frac{k-k_{1}-k_{2}}{2}+\hat{q}\right)^{2}-m^{2}_{c}}
\slashed{\epsilon}^{\ast}(k_{1})\frac{\frac{\slashed{k}
+\slashed{k}_{1}-\slashed{k}_{2}}{2}+\hat{\slashed{q}}+m_{c}}
{\left(\frac{k+k_{1}-k_{2}}{2}+\hat{q}\right)^{2}-m^{2}_{c}}\slashed{\epsilon}^{\ast}(k_{2})
\Bigg{]}
\end{eqnarray}
with the $c$ quark mass $m_{c}$.

\begin{figure}[h]
\centering
\includegraphics[width=0.45\textwidth]{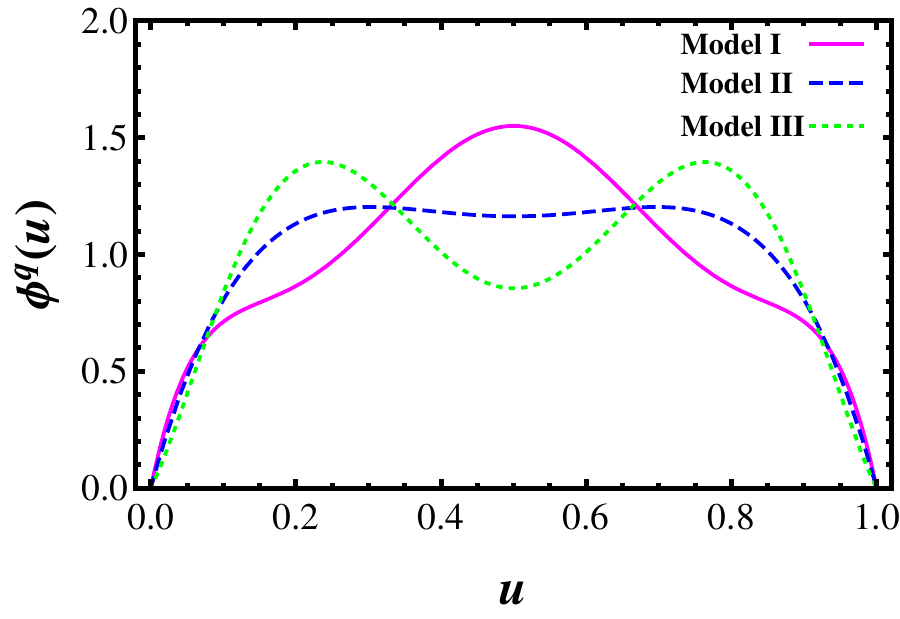}
\caption{The shapes of the corresponding DAs at the scale of $\mu=m_{c}$.}\label{DAs}
\end{figure}

The light-cone expansion of the matrix elements of $\eta^{(\prime)}$ over quark and antiquark fields reads~\cite{Chernyak:1983ej,Ali:2003kg,Ball:2007hb}
\begin{eqnarray}
\langle \eta^{(\prime)}(p) |\bar{q}_{\alpha}(x)q_{\beta}(y)|0\rangle &=&\frac{i}{4}f_{\eta^{(\prime)}}^{q}\left(\slashed{p}\gamma_{5}\right)_{\beta\alpha}
 \int\textrm{d}u e^{i(\bar{u}p\cdot y+up\cdot x)}\phi^{q}(u)+\cdots,
\end{eqnarray}
where $p$ represents the momentum of $\eta^{(\prime)}$, the superscript $q=u,d,s$ denotes the flavor of the light quarks, the ``$\cdots$" stands for the high twist terms and the decay constants $f_{\eta^{(\prime)}}^{q}$ are defined as
\begin{eqnarray}
\langle0|\bar{q}(0)\gamma_{\mu}\gamma_{5}q(0)|\eta^{(\prime)}(p)\rangle&=&
i f_{\eta^{(\prime)}}^{q}p_{\mu}.
\end{eqnarray}
Then one can obtain the coupling of $g^{\ast}g^{\ast}-\eta^{(\prime)}$ up to twist-3 level~\cite{Muta:1999tc,Yang:2000ce,Ali:2000ci}:
\begin{eqnarray}
\mathcal{M}^{\mu\nu}\epsilon_{\mu}(k_{1})\epsilon_{\nu}(k_{2})&=&-i (4\pi \alpha_{s})\delta_{ab}\epsilon^{\mu\nu\rho\sigma}\epsilon_{\mu}(k_{1})\epsilon_{\nu}(k_{2})k_{1\rho}k_{2\sigma}\nonumber \\
& &\times\sum_{q=u,d,s}\frac{f_{\eta^{(\prime)}}^{q}}{6}
\int^{1}_{0}du\phi^{q}(u)\left(\frac{1}{\bar{u}k_{1}^{2}+uk_{2}^{2}-u\bar{u}p^{2}-m_{q}^{2}}+(u\leftrightarrow\bar{u})\right)
\end{eqnarray}
with $\bar{u}=1-u$. Here $u$ is the momentum fraction carried by the quark, $m_{q}$ is the mass of the quark ($q=u,d,s$).  The light-cone DA has the form~\cite{Agaev:2014wna}
\begin{eqnarray}
\phi^{q}(u)&=&\phi_{AS}(u)\left[1+\sum_{n=2,4\cdots}c^{q}_{n}(\mu)C_{n}^{\frac{3}{2}}(2u-1)\right]
\end{eqnarray}
with the asymptotic form of DA $\phi_{AS}(u)=6u(1-u)$ and the Gegenbauer moments $c^{q}_{n}(\mu)$.
In Table~\ref{tab:coefficients}, we list three models of the DAs given in Ref.~\cite{Agaev:2014wna}. Schematically, we also show their shapes at the scale of $\mu_{0}=m_{c}$ in Fig.~\ref{DAs}. As pointed out in Refs.~\cite{He:2019mpy,Fan:2019sap}, the decay rates of $h_{c}\rightarrow\gamma\eta^{(\prime)}$ barely depend on the shapes of $\eta^{(\prime)}$ DAs (we will estimate them below). It means that the mixing angle of $\eta-\eta^{\prime}$ system could be reliably extracted in our calculations due to the negligible uncertainties from $\eta^{(\prime)}$ DAs.
\begin{table}[h]
  \caption{Gegenbauer moments of three sample models at the scale of $\mu_{0}=1\, \mathrm{GeV}$.}\label{tab:coefficients}
\vspace{0.2cm}
\centering
  \begin{tabular}{l c r @{.} l c}
  \hline\hline
   Model    &~~~~~~~~~~~~~~$c_{2}^{q}(\mu_{0})$~~~~~~~~~~~~~~&\multicolumn{2}{c}{$c_{4}^{q}(\mu_{0})$}
            ~~~~~~~~~~~~~~&$c_{2}^{g}(\mu_{0})$    \\
  \hline
  I            &   $0.10$   &   $0$&$10$    &   $-0.26$ \\
  II           &   $0.20$   &   $0$&$00$    &   $-0.31$ \\
  III          &   $0.25$   &   $-0$&$10$   &   $-0.25$ \\
  \hline\hline
  \end{tabular}
\end{table}

To proceed, the decay amplitude of $h_{c}\rightarrow\gamma\eta^{(\prime)}$ can be obtained directly by contracting the two couplings ${\mathcal A}^{\alpha\beta\mu\nu}$ and $\mathcal{M}_{\mu\nu}$, inserting the gluon propagators and integrating over the loop momentum:
\begin{equation}
M_{T}=T^{\alpha\beta}\varepsilon_{\alpha}(K)\epsilon^{\ast}_{\beta}(k)=\frac{1}{2}\int\frac{\mathrm{d}^{4}k_{1}}{(2\pi)^{4}}{\mathcal A}^{\alpha\beta\mu\nu}\mathcal{M}_{\mu\nu}\frac{i}{k^{2}_{1}
+i\epsilon}\frac{i}{k^{2}_{2}+i\epsilon}\varepsilon_{\alpha}(K)\epsilon^{\ast}_{\beta}(k),
\end{equation}
where the factor $1/2$ takes into account that the two gluons have already been interchanged in both $A^{\alpha\beta\mu\nu}$ and $M_{\mu\nu}$. By Lorentz invariance, parity conservation and gauge invariance, one can obtain~\cite{Korner:1982vg}
\begin{equation}
T^{\alpha\beta}\propto  -g^{\alpha\beta}+\frac{k^{\alpha}K^{\beta}}{K\cdot k},
\end{equation}
i.e., there is only one independent helicity amplitude $H_{QCD}^{q}$:
\begin{eqnarray}\label{Tee}
T^{\alpha\beta}\varepsilon_{\alpha}(K)\epsilon^{\ast}_{\beta}(k)
&=&H_{QCD}^{q}h^{\alpha\beta}\varepsilon_{\alpha}(K)\epsilon^{\ast}_{\beta}(k),
\end{eqnarray}
where
\begin{eqnarray}
h^{\alpha\beta}=-g^{\alpha\beta}+\frac{k^{\alpha}K^{\beta}}{k\cdot K}.
\end{eqnarray}
With the help of the helicity projector~\cite{Korner:1982vg}
\begin{eqnarray}
\mathbb{P}^{\alpha\beta}=\frac{1}{2}h^{\ast}_{\alpha^{\prime}\beta^{\prime}}\left(-g^{\alpha\alpha^{\prime}}
+\frac{K^{\alpha}K^{\alpha^{\prime}}}{M^{2}}\right)\left(-g^{\beta\beta^{\prime}}\right)=\frac{1}{2}\left(-g^{\alpha\beta}+
\frac{k^{\alpha}K^{\beta}}{k\cdot K}\right),
\end{eqnarray}
one can obtain the helicity amplitude
\begin{eqnarray}\label{scalarq}
H_{QCD}^{q}=T^{\alpha\beta}\mathbb{P}_{\alpha\beta}=\frac{2Q_{c}}{3\sqrt{3}}\sqrt{4\pi\alpha}(4\pi\alpha_{s})^{2}
\sum_{q=u,d,s}f_{\eta^{(\prime)}}^{q}H_{q},
\end{eqnarray}
where the dimensionless function $H_{q}$ reads
\begin{eqnarray}\label{dlesshq}
 H_{q}
&=&\int\frac{\mathrm{d}^{3}\hat{q}}{(2\pi)^{3}}f(\hat{q}^{2})
\int\textrm{d}u\phi^{q}(u)I_{q}(u,\hat{q}).
\end{eqnarray}
$I_{q}(u,\hat{q})$ represents the sum of the loop integrals of all the Feynman diagrams
\begin{eqnarray}\label{loopfunctionI}
I_{q}(u,\hat{q})=\int\frac{\textrm{d}^{4}l}{(2\pi)^{4}}
\Bigg{(}\frac{N_{1}}{D_{1}D_{2}D_{3} D_{4}D_{5}}+\frac{N_{2}}{C_{1}D_{1} D_{3} D_{4} D_{5}}+\frac{N_{3}}{C_{2}D_{1}D_{2} D_{4} D_{5}}\Bigg{)}+(u\leftrightarrow\bar{u})
\end{eqnarray}
with $l=k_{1}-k_{2}$ and the denominators of the propagators
\begin{eqnarray}
C_{1}&=&(p-k+2\hat{q})^{2}-4m_{c}^{2}+i\epsilon,\nonumber\\
C_{2}&=&(p-k-2\hat{q})^{2}-4m_{c}^{2}+i\epsilon,\nonumber\\
D_{1}&=&\left[l+(\bar{u}-u) p\right]^{2}-4m_{q}^{2}+i\epsilon,\nonumber\\
D_{2}&=&(l-k-2\hat{q})^{2}-4m_{c}^{2}+i\epsilon,\nonumber\\
D_{3}&=&(l+k-2\hat{q})^{2}-4m_{c}^{2}+i\epsilon,\nonumber\\
D_{4}&=&(l+p)^{2}+i\epsilon,\nonumber\\
D_{5}&=&(l-p)^{2}+i\epsilon.
\end{eqnarray}
As shown in Eq.~\eqref{dlesshq}, the spin structures of the bound-state wave function are absorbed into the loop function $I_{q}(u,\hat{q})$, and the expressions of the numerators $N_{1}$, $N_{2}$ and $N_{3}$ are presented in the Appendix. Since the loop function $I_{q}(u,\hat{q})$
has no soft singularities and the dimensionless function $H_{q}$ is very insensitive to the light quark mass $m_{q}$~\cite{He:2019mpy,Fan:2019sap}, one can take the following simplicity safely:
\begin{eqnarray}\label{scalar0}
I_{0}(u,\hat{q})&=&\lim_{m_{q}\to 0} I_{q}(u,\hat{q}),\\\nonumber
H_{0}&=&\int\frac{\mathrm{d}^{3}\hat{q}}{(2\pi)^{3}}f(\hat{q}^{2})
\int\textrm{d}u\phi^{q}(u)I_{0}(u,\hat{q}),
\end{eqnarray}
i.e., $H_{q}(q=u,d,s)= H_{0}$. Then the helicity amplitude in Eq.~\eqref{scalarq} can be rewritten as
\begin{eqnarray}
H_{QCD}^{q}&=&\frac{2Q_{c}}{3\sqrt{3}}\sqrt{4\pi\alpha}(4\pi\alpha_{s})^{2}f_{\eta^{(\prime)}}H_{0}
\end{eqnarray}
with the effective decay constants
\begin{eqnarray}
f_{\eta^{\prime}}=f_{\eta^{\prime}}^{u}+f_{\eta^{\prime}}^{d}+f_{\eta^{\prime}}^{s},\quad\quad
f_{\eta}=f_{\eta}^{u}+f_{\eta}^{d}+f_{\eta}^{s}.
\end{eqnarray}

By using the algebraic identity~($\xi\neq\pm 1$)
\begin{equation}
\frac{1}{m^{2}(\xi^{2}-1)}D_{1}-\frac{1}{2m^{2}(\xi-1)}D_{4}+\frac{1}
{2m^{2}(\xi+1)}D_{5}=1
\end{equation}
with $\xi=1-2u$ and the mass of $\eta^{(\prime)}$ meson $m$, the loop function $I_{0}(u,\hat{q})$ can be decomposed into a sum of four-point one-loop integrals
\begin{eqnarray}
I_{0}(u,\hat{q})&=&\int\frac{\textrm{d}^{4}l}{(2\pi)^{4}}
\Bigg{(}\frac{N_{1}}{m^{2}(\xi^{2}-1)D_{2}D_{3} D_{4}D_{5}}-\frac{N_{1}}{2m^{2}(\xi-1)D_{1}D_{2}D_{3} D_{5}}\nonumber\\
&+&\frac{N_{1}}{2m^{2}(\xi+1)D_{1}D_{2}D_{3} D_{4}}+\frac{N_{2}}{C_{1}D_{1} D_{3} D_{4} D_{5}}+\frac{N_{3}}{C_{2}D_{1}D_{2} D_{4} D_{5}}\Bigg{)}+(u\leftrightarrow\bar{u}).
\end{eqnarray}
When $\xi=1$, the denominators of the propagators have the relation $D_{1}=D_{4}$,
and the loop function $I_{0}(u,\hat{q})$ becomes
\begin{eqnarray}
I_{0}(u,\hat{q})&=&\int\frac{\textrm{d}^{4}l}{(2\pi)^{4}}
\Bigg{[}\frac{N_{1}}{D_{2}D_{3} D^{2}_{4}D_{5}}+\frac{N_{2}}{C_{1} D_{3} D^{2}_{4} D_{5}}+\frac{N_{3}}{C_{2}D_{2} D^{2}_{4} D_{5}}\Bigg{]}+(u\leftrightarrow\bar{u}).
\end{eqnarray}
And when $\xi=-1$, the denominators of the propagators have the relation $D_{1}=D_{5}$,
then the loop function $I_{0}(u,\hat{q})$ becomes
\begin{eqnarray}
I_{0}(u,\hat{q})=\int\frac{\textrm{d}^{4}l}{(2\pi)^{4}}
\Bigg{[}\frac{N_{1}}{D_{2}D_{3} D_{4}D^{2}_{5}}+\frac{N_{2}}{C_{1} D_{3} D_{4} D^{2}_{5}}+\frac{N_{3}}{C_{2}D_{2} D_{4} D^{2}_{5}}\Bigg{]}+(u\leftrightarrow\bar{u}).
\end{eqnarray}
With the program $Package-\mathrm{X}$~\cite{Patel:2015tea,Patel:2016fam}, one can evaluate the above one-loop integrals analytically. Similar to the situations without considering the internal momentum of charmonium~\cite{He:2019mpy,Fan:2019sap}, we find that the loop function $I_{0}(u,\hat{q})$ is also almost unchanged over the most region of the momentum fraction $u$, and this results in the dimensionless function $H_{0}$ very insensitive to the shapes of the $\eta^{(\prime)}$ DAs. Numerically, our results show that the change among the dimensionless function $H_{0}$ with the different models of the DAs in Fig.~\ref{DAs} is less than $1\%$. Therefore, the theoretical uncertainties from the DAs are ignorable in our calculations of the branching ratios $\mathcal{B}(h_c\rightarrow\gamma\eta^{(\prime)})$. In addition, due to the fact that the internal momentum $\hat{q}$ could make the propagator on-shell in the region of the wave function of $h_{c}$ unsuppressed strongly, the convolution of the loop function $I_{0}(u,\hat{q})$ and the wave function $f(\hat{q}^{2})$ (i.e., the dimensionless function $H_{0}$) would gain substantial kinematical corrections. Specifically, there is a significant enhancement in the absorptive part of $H_{0}$. Accordingly, the relativistic effects are important in the quark-antiquark contributions.

\subsection{The contributions of the gluonic content of $\eta^{(\prime)}$}
\label{subsec:QCDg}
\begin{figure}[th]
  \begin{center}
  \includegraphics[width=0.45\textwidth]{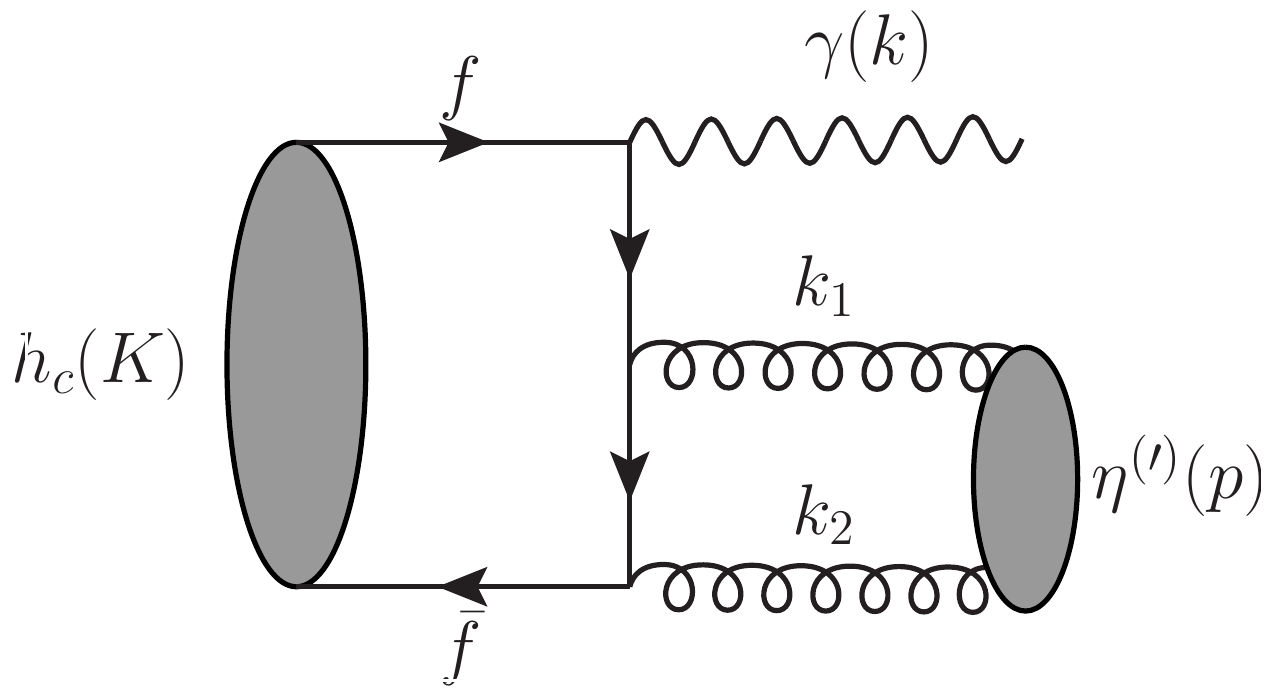}
  \end{center}
  \vskip -0.7cm
  \caption{One typical Feynman diagram for $h_{c}\rightarrow\gamma\eta^{(\prime)}$ with the gluonic content of $\eta^{(\prime)}$. Here the kinematic variables are labeled.}\label{hQCDg}
\end{figure}

The gluonic content of $\eta^{(\prime)}$ can directly contribute to the decay processes $h_{c}\rightarrow\gamma\eta^{(\prime)}$ from tree level. One typical Feynman diagram is exhibited in Fig.~\ref{hQCDg}, and there are other two diagrams from permutations of the photon and the gluon legs. Generally, contributions of gluonic content are supposed to be small~\cite{Escribano:2007cd,He:2019mpy}, since gluonic content can be seen as the higher-order effects from the point of view of the QCD evolution of the two-gluon DA, which vanishes in the asymptotic limit. However, as we have pionted out in Ref.~\cite{Fan:2019sap}, these contributions may become important in the $\eta^{(\prime)}$ production because the two-gluon DA of $\eta^{(\prime)}$ can mix with their quark-antiquark DA due to the $U(1)_{A}$ anomaly. Furthermore, from Figs.~\ref{hQCDq} and~\ref{hQCDg}, one can easily find that the leading-order contributions (one-loop level) from the quark-antiquark content of $\eta^{(\prime)}$ are suppressed by a factor of $\alpha_{s}$ as compared with the contributions from the gluonic content of $\eta^{(\prime)}$. Therefore, there is an interesting question: which kind of contributions is dominant in the decays $h_{c}\rightarrow\gamma\eta^{(\prime)}$, especially with considering the relativistic effects in the two decay processes, and the answer is given below.

The matrix elements of the mesons $\eta^{(\prime)}$
over two-gluon fields in the light-cone expansion at the leading-twist level read~\cite{Kroll:2002nt,Ball:2007hb,Agaev:2014wna}:
\begin{eqnarray}
\langle\eta^{(\prime)}(p)|A_{\alpha}^{a}(x)A_{\beta}^{b}(y)|0\rangle=\frac{1}{4}\epsilon_{\alpha\beta\mu\nu}
\frac{k^{\mu}p^{\nu}}{p\cdot k}\frac{C_{F}}{\sqrt{3}}\frac{\delta^{ab}}{8}f_{\eta^{(\prime)}}^{1}\int\textrm{d}u e^{i(up\cdot x+\bar{u}p\cdot y)}\frac{\phi^{g}(u)}{u(1-u)}
\end{eqnarray}
with the effective decay constant $f_{\eta^{(\prime)}}^{1}=\frac{1}{\sqrt{3}}(f_{\eta^{(\prime)}}^{u}+f_{\eta^{(\prime)}}^{d}+f_{\eta^{(\prime)}}^{s})$ and the gluonic twist-2 DA~\cite{Agaev:2014wna,Ball:2007hb,Alte:2015dpo}
\begin{eqnarray}
\phi^{g}(u)=30u^{2}(1-u)^{2}\sum_{n=2,4\cdots}c^{g}_{n}(\mu)C_{n-1}^{\frac{5}{2}}(2u-1).
\end{eqnarray}
One can obtain the corresponding helicity amplitude
\begin{eqnarray}
H_{QCD}^{g}&=&\frac{2Q_{c}}{9} \sqrt{4\pi\alpha}(4\pi\alpha_{s})f_{\eta^{(\prime)}}^{1} H_{g},
\end{eqnarray}
where the dimensionless function $H_{g}$ has the form
\begin{eqnarray}\label{Hg}
H_{g}&=&\int\frac{\mathrm{d}^{3}\hat{q}}{(2\pi)^{3}}f(\hat{q}^{2})\int\mathrm{d}u\frac{\phi^{g}(u)}{u(1-u)}
\left(\frac{N_{4}}{C_{1}C_{4}}+\frac{N_{5}}{C_{2}C_{3}}+\frac{N_{6}}{C_{3}C_{4}}\right).
\end{eqnarray}
Here the denominators of the propagators $C_{3}$ and $C_{4}$ read
\begin{eqnarray}
C_{3}&=&(\xi p+k+2\hat{q})^{2}-4m_{c}^{2}+i\epsilon,\nonumber\\
C_{4}&=&(\xi p-k+2\hat{q})^{2}-4m_{c}^{2}+i\epsilon,
\end{eqnarray}
and the expressions of the numerators $N_{4}$, $N_{5}$ and $N_{6}$ are given in the Appendix.

In the remaining part of this section, we present a brief discussion about the relativistic effects. To the next-to-leading-order corrections related to the internal momentum from the numerators $N_{4}$, $N_{5}$ and $N_{6}$, if we take $\hat{m}_{c}\approx m_{c}\approx M/2$ and $m^{2}/M^{2}\approx 0$, they exhibit the following behavior:
\begin{eqnarray}\label{threenumerators}
N_{4}&\propto&\left(1-\frac{\xi}{1+\xi}\frac{\hat{q}^{2}}{k\cdot\hat{q}}+{\mathcal O}(\hat{q}^{2})\right),\nonumber\\
N_{5}&\propto&\left(1-\frac{\xi}{1-\xi}\frac{\hat{q}^{2}}{k\cdot\hat{q}}+{\mathcal O}(\hat{q}^{2})\right),\nonumber\\
N_{6}&\propto&\left(1+\frac{\xi}{1-\xi^{2}}\frac{\hat{q}^{2}}{k\cdot\hat{q}}+{\mathcal O}(\hat{q}^{2})\right).
\end{eqnarray}
Obviously, the next-to-leading-order contributions are not suppressed enough in the major region\renewcommand{\thefootnote}{\fnsymbol{footnote}}\footnote[2]{Of course, the contributions from the large integration variable $\hat{q}$ would be strongly suppressed by the bound-state wave function (see Eqs.~\eqref{Hg} and~\eqref{fhatq}). Empirically, the major region of the wave function of the charmonia may be supposed near or below $1~\mathrm{GeV}$~\cite{Bhatnagar:2016otj}.} of the integration variable $\hat{q}$. Therefore, in the decay processes $h_{c}\rightarrow\gamma\eta^{(\prime)}$, the relativistic corrections should be taken into account.

\section{Numerical results}
\label{sec:numerical analysis}
The decay widths of $h_{c}\rightarrow \gamma\eta^{(\prime)}$ can be expressed as
\begin{eqnarray}
\Gamma(h_{c}\rightarrow \gamma\eta^{(\prime)})=\frac{2}{3}\frac{1-x}{16\pi M}\left|H_{QCD}^{q}+H_{QCD}^{g}\right|^{2}
\end{eqnarray}
with $x=m^{2}/M^{2}$. In the following numerical calculations, we take the parameters: $M=3525~\mathrm{MeV}$, $m_{\eta}=548~\mathrm{MeV}$, $m_{\eta^{\prime}}= 958~\mathrm{MeV}$, $m_{c}= 1270~\mathrm{MeV}$, $\Gamma_{h_c}=(0.70\pm0.28\pm0.22)~\mathrm{MeV}$ and $f_{\pi}=130.2~\mathrm{MeV}$, which are quoted from the PDG~\cite{Zyla:2020zbs}. The QCD running coupling constant is adopted $\alpha_{s}(m_{c})=0.38$, which is calculated through the two-loop renormalization group equation. The effective mass of $c$ quark and the harmonic oscillator parameter appearing in the bound-state wave function, which contains the long-distance nonperturbative dynamical effect of quark-antiquark interaction, are respectively taken as $\hat{m}_{c}=1490~\mathrm{MeV}$ and $\beta_{A}=590~\mathrm{MeV}$, and more discussions could be found in Refs~\cite{Negash:2015rua,Bhatnagar:2016otj,Gebrehana:2019mpw}. As we have already mentioned, the theoretical uncertainties from $\eta^{(\prime)}$ DAs are negligible. So in our calculations, we choose the Model I of the meson DA in Table~\ref{tab:coefficients}.

For the mixing of $\eta-\eta^{\prime}$ system, we take the single-mixing-angle scheme in quark-flavor basis~\cite{Akhoury:1987ed,Ball:1995zv,Feldmann:1998vh,Feldmann:1998sh,Feldmann:1999uf}, and then the effective decay constants can be parameterized as
\begin{eqnarray}
 f^{u(d)}_{\eta} &=& \frac{f_{q}}{\sqrt{2}}\cos\phi,~~~~~~~~~~ f^{s}_{\eta}=-f_{s}\sin\phi,\nonumber\\
f^{u(d)}_{\eta^{\prime}} &=& \frac{f_{q}}{\sqrt{2}}\sin\phi,~~~~~~~~~~   f^{s}_{\eta^{\prime}}=f_{s}\cos\phi.
\end{eqnarray}
Here the mixing angle $\phi$ and the decay constants $f_{q}$, $f_{s}$ are three phenomenological parameters
which can been determined in
different methods (see Refs.~\cite{Feldmann:1998vh,Escribano:2007cd,Michael:2013gka,Escribano:2013kba,Chen:2014yta,
Guo:2015xva,Escribano:2013kba,Escribano:2015nra,Escribano:2015yup,Ottnad:2017bjt,Gan:2020aco,Escribano:2020jdy} and references therein).

\subsection{Branching ratios}
\label{subsec:branchingratios}

The phenomenological parameters $\phi$, $f_{q}$ and $f_{s}$ have been determined in~\cite{Feldmann:1998vh} as
\begin{eqnarray}
\phi=39.3^{\circ}\pm 1.0^{\circ},~~~~f_{q}=(1.07\pm 0.02)f_{\pi},~~~~f_{s}=(1.34\pm 0.06)f_{\pi},
\end{eqnarray}
which are the known FKS results. With the set of parameter values and $\Gamma_{h_c} = 0.70\, \mathrm{MeV}$, we obtain our numerical results of the contributions from different contents of $\eta^{(\prime)}$ in Tables~\ref{tab:FKSQCDq},~\ref{tab:FKSQCDg} and~\ref{tab:FKStotal}, where the branching ratios $\mathcal{B}(h_{c}\rightarrow\gamma\eta)$, $\mathcal{B}(h_{c}\rightarrow\gamma\eta^{\prime})$ and their ratio $R_{h_{c}}=\mathcal{B}(h_{c}\rightarrow\gamma\eta)/\mathcal{B}(h_{c}\rightarrow\gamma\eta^{\prime})$ are presented in the first, second and third lines of these tables, respectively. The contributions from the quark-antiquark content of $\eta^{(\prime)}$ and those from the gluonic content of $\eta^{(\prime)}$ are presented in Tables~\ref{tab:FKSQCDq} and~\ref{tab:FKSQCDg}, respectively. The total contributions both from the quark-antiquark content and the gluonic content of $\eta^{(\prime)}$ are presented in Table~\ref{tab:FKStotal}. In order to show the contributions from the relativistic effects more clearly, we present the results with the zero-binding approximation\renewcommand{\thefootnote}{\fnsymbol{footnote}}\footnote[2]{We update the previous results~\cite{Fan:2019sap} with the QCD running coupling constant $\alpha_{s}(m_{c})=0.38$.} (i.e., without relativistic corrections) ~\cite{Fan:2019sap} and those with the relativistic corrections in the first and second columns of these tables respectively.

\begin{table}[!htbp]
  \caption{\label{tab:FKSQCDq}The quark-antiquark contributions obtained with the zero-binding approximation and the B-S formalism, respectively.}
  \vspace{0.2cm}
  \centering
  \begin{tabular}{lcccc}
  \hline\hline
   &  zero-binding~\cite{Fan:2019sap}   &  this work  & Exp.~\cite{Ablikim:2016uoc}    \\
  \hline
 $\mathcal{B}(h_{c}\rightarrow\gamma\eta)$~~&~~$0.1\times10^{-4}$~~&~~$0.3\times10^{-4}$~~&~~$(4.7\pm1.5\pm1.4)\times10^{-4}$ \\
 $\mathcal{B}(h_{c}\rightarrow\gamma\eta^{\prime})$~~&~~$0.38\times10^{-3}$~~&~~$0.87\times10^{-3}$~~&~~$(1.52\pm0.27\pm0.29)\times10^{-3}$\\
 $R_{h_{c}}$                ~~&~~$3.4\%$~~&~~$4.0\%$ ~~&~~$(30.7\pm11.3\pm8.7)\%$  \\
  \hline\hline
  \end{tabular}
\end{table}
\begin{table}[!htbp]
  \caption{\label{tab:FKSQCDg}The gluonic contributions obtained with the zero-binding approximation and the B-S formalism, respectively.}
  \vspace{0.2cm}
  \centering
  \begin{tabular}{lcccc}
  \hline\hline
   &  zero-binding~\cite{Fan:2019sap}   &  this work  & Exp.~\cite{Ablikim:2016uoc}    \\
  \hline
 $\mathcal{B}(h_{c}\rightarrow\gamma\eta)$~~&~~$0.1\times10^{-4}$~~&~~$0.1\times10^{-4}$~~&~~$(4.7\pm1.5\pm1.4)\times10^{-4}$ \\
 $\mathcal{B}(h_{c}\rightarrow\gamma\eta^{\prime})$~~&~~$0.27\times10^{-3}$~~&~~$0.45\times10^{-3}$~~&~~$(1.52\pm0.27\pm0.29)\times10^{-3}$\\
 $R_{h_{c}}$                ~~&~~$3.0\%$~~&~~$2.9\%$ ~~&~~$(30.7\pm11.3\pm8.7)\%$  \\
  \hline\hline
  \end{tabular}
\end{table}
\begin{table}[!htbp]
  \caption{\label{tab:FKStotal}Both the quark-antiquark and gluonic contributions obtained with the zero-binding approximation and the B-S formalism, respectively.}
  \vspace{0.2cm}
  \centering
  \begin{tabular}{lcccc}
  \hline\hline
   &  zero-binding~\cite{Fan:2019sap}   &  this work  & Exp.~\cite{Ablikim:2016uoc}    \\
  \hline
 $\mathcal{B}(h_{c}\rightarrow\gamma\eta)$~~&~~$0.3\times10^{-4}$~~&~~$0.9\times10^{-4}$~~&~~$(4.7\pm1.5\pm1.4)\times10^{-4}$ \\
 $\mathcal{B}(h_{c}\rightarrow\gamma\eta^{\prime})$~~&~~$0.92\times10^{-3}$~~&~~$2.29\times10^{-3}$~~&~~$(1.52\pm0.27\pm0.29)\times10^{-3}$\\
 $R_{h_{c}}$                ~~&~~$3.8\%$~~&~~$3.8\%$ ~~&~~$(30.7\pm11.3\pm8.7)\%$  \\
  \hline\hline
  \end{tabular}
\end{table}

From the Tables~\ref{tab:FKSQCDq} and~\ref{tab:FKSQCDg}, one can find that the gluonic contributions and the quark-antiquark contributions are comparable with each other, whatever the relativistic corrections are taken into account or not. It is unlike the situation where the gluonic contributions are strongly suppressed in the heavy vector quarkonium decays $V\rightarrow \gamma\eta^{(\prime)}$~\cite{Baier:1981pm,Ma:2002ww,He:2019mpy} because of an additional suppression factor (i.e., $m^{2}/M^{2}$) from the spin structure of their amplitudes. Besides, our results are also unlike the phenomenological fits where the gluonic content can be neglected~\cite{Escribano:2007cd,Escribano:2008rq,Ambrosino:2009sc}, especially for the meson $\eta$. By comparing with the results of the zero-binding approximation, we find out that the relativistic corrections are significant for both the quark-antiquark contributions and the gluonic contributions. Intriguingly, the importance of the gluonic contributions in the decay processes $h_{c}\rightarrow\gamma\eta^{(\prime)}$ indicates that these two decay processes can test the gluonic content of $\eta^{(\prime)}$ more efficiently than the decay processes $V\rightarrow \gamma\eta^{(\prime)}$.

Comparing the results listed in Tables~\ref{tab:FKSQCDq} and~\ref{tab:FKSQCDg} with those listed in Table~\ref{tab:FKStotal}, we find that the branching ratios
both $\mathcal{B}(h_{c}\rightarrow\gamma\eta)$ and $\mathcal{B}(h_{c}\rightarrow\gamma\eta^{\prime})$ are greatly enhanced with the constructive interference of the quark-antiquark contributions and the gluonic contributions, whatever the relativistic corrections are taken into account or not. Unfortunately, the branching ratio $\mathcal{B}(h_{c}\rightarrow\gamma\eta)$ is still much smaller than its experimental value~\cite{Ablikim:2016uoc}, even though it is substantially enhanced by the relativistic corrections. Moreover, it is thought-provoking that the ratio $R_{h_{c}}$ with containing more dynamical corrections from the initial meson $h_{c}$ hardly change\renewcommand{\thefootnote}{\fnsymbol{footnote}}\footnote[2]{The main reason is that there is no node in the wave function of $h_{c}$~\cite{Silverman:1987de,Beyer:1992nd,Bhatnagar:2016otj}, so the relativistic corrections at large extent could cancel in the ratio $R_{h_{c}}$.} and is still smaller than the experimental value~\cite{Ablikim:2016uoc}. These might imply the set of parameters of FKS results is not unquestionable.

As we have already mentioned, there are some obvious discrepancies in the determinations of the mixing angle~\cite{Feldmann:1998vh,Gregory:2011sg,Michael:2013gka,Escribano:2013kba,
Chen:2014yta,Escribano:2015nra,Escribano:2015yup,Urbach:2017rvx,He:2019mpy}. Besides the value of the mixing angle $\phi \sim 40^{\circ}$ (see, e.g., Refs~\cite{Feldmann:1998vh,Michael:2013gka,Escribano:2013kba,
Escribano:2015nra,Escribano:2015yup,Urbach:2017rvx}), a smaller value of the mixing angle $\phi \sim 34^{\circ}$ is also usually obtained in many methods~\cite{Gregory:2011sg,Escribano:2013kba,
Chen:2014yta,He:2019mpy}. Therefore, it would be interesting to show the results of the branching ratios $\mathcal{B}(h_{c}\rightarrow\gamma\eta)$, $\mathcal{B}(h_{c}\rightarrow\gamma\eta^{\prime})$ and their ratio $R_{h_{c}}$ with the different sets of the phenomenological parameters.

With the set of the parameter values~\cite{Escribano:2013kba}
\begin{eqnarray}
\phi=33.5^{\circ}\pm 0.9^{\circ},~~~~f_{q}=(1.09\pm 0.02)f_{\pi},~~~~f_{s}=(0.96\pm 0.04)f_{\pi},
\end{eqnarray}
extracted from the transition form factor (TFF) $F_{\gamma^{\ast}\gamma\eta^{\prime}}(+\infty)$, which is in accord with the BABAR measurement in the timelike region at $q^{2}=112\, \mathrm{GeV}^{2}$~\cite{Aubert:2006cy}, we present the numerical results in Tables~\ref{tab:TFFQCDq},~\ref{tab:TFFQCDg} and~\ref{tab:TFFtotal} with $\Gamma_{h_c} = 0.70\, \mathrm{MeV}$.

\begin{table}[!htbp]
  \caption{\label{tab:TFFQCDq}The quark-antiquark contributions obtained with a smaller value of the mixing angle.}
  \vspace{0.2cm}
  \centering
  \begin{tabular}{lcccc}
  \hline\hline
 ~~&~~zero-binding~\cite{Fan:2019sap}~~~ &~~~this work~~&~~Exp.~\cite{Ablikim:2016uoc}~~~  \\
  \hline
 $\mathcal{B}(h_{c}\rightarrow\gamma\eta)$~~&~~$0.7\times10^{-4}$~~&~~$1.9\times10^{-4}$~~&~~$(4.7\pm1.5\pm1.4)\times10^{-4}$ \\
 $\mathcal{B}(h_{c}\rightarrow\gamma\eta^{\prime})$~~&~~$0.26\times10^{-3}$~~&~~$0.6\times10^{-3}$~~&~~$(1.52\pm0.27\pm0.29)\times10^{-3}$\\
 $R_{h_{c}}$                ~~&~~$27.5\%$~~&~~$31.7\%$ ~~&~~$(30.7\pm11.3\pm8.7)\%$  \\
  \hline\hline
  \end{tabular}
\end{table}
\begin{table}[!htbp]
  \caption{\label{tab:TFFQCDg}The gluonic contributions obtained with a smaller value of the mixing angle.}
  \vspace{0.2cm}
  \centering
  \begin{tabular}{lcccc}
  \hline\hline
 ~~&~~zero-binding~\cite{Fan:2019sap}~~~ &~~~this work~~&~~Exp.~\cite{Ablikim:2016uoc}~~~  \\
  \hline
 $\mathcal{B}(h_{c}\rightarrow\gamma\eta)$~~&~~$0.4\times10^{-4}$~~&~~$0.7\times10^{-4}$~~&~~$(4.7\pm1.5\pm1.4)\times10^{-4}$ \\
 $\mathcal{B}(h_{c}\rightarrow\gamma\eta^{\prime})$~~&~~$0.19\times10^{-3}$~~&~~$0.31\times10^{-3}$~~&~~$(1.52\pm0.27\pm0.29)\times10^{-3}$\\
 $R_{h_{c}}$                ~~&~~$23.8\%$~~&~~$23.4\%$ ~~&~~$(30.7\pm11.3\pm8.7)\%$  \\
  \hline\hline
  \end{tabular}
\end{table}
\begin{table}[!htbp]
  \caption{\label{tab:TFFtotal}Both the quark-antiquark and gluonic contributions obtained with a smaller value of the mixing angle.}
  \vspace{0.2cm}
  \centering
  \begin{tabular}{lcccc}
  \hline\hline
 ~~&~~zero-binding~\cite{Fan:2019sap}~~~ &~~~this work~~&~~Exp.~\cite{Ablikim:2016uoc}~~~  \\
  \hline
 $\mathcal{B}(h_{c}\rightarrow\gamma\eta)$~~&~~$1.9\times10^{-4}$~~&~~$4.7\times10^{-4}$~~&~~$(4.7\pm1.5\pm1.4)\times10^{-4}$ \\
 $\mathcal{B}(h_{c}\rightarrow\gamma\eta^{\prime})$~~&~~$0.63\times10^{-3}$~~&~~$1.57\times10^{-3}$~~&~~$(1.52\pm0.27\pm0.29)\times10^{-3}$\\
 $R_{h_{c}}$                ~~&~~$30.2\%$~~&~~$30.3\%$ ~~&~~$(30.7\pm11.3\pm8.7)\%$  \\
  \hline\hline
  \end{tabular}
\end{table}

As we have already mentioned, theoretical calculation of the branching ratio $\mathcal{B}(h_{c}\rightarrow\gamma\eta)$ is certainly sensitive to the mixing angle of $\eta-\eta^{\prime}$ mixing, and the main reason is that the decay amplitude is proportional to the factor $(\sqrt{2}f_{q}\cos\phi-f_{s}\sin\phi)$, which lead to large cancellations in the matrix elements. More interestingly, after taking into account the relativistic corrections, not only the ratio $R_{h_{c}}$ but also the individual branching ratios $\mathcal{B}(h_{c}\rightarrow\gamma\eta)$ and $\mathcal{B}(h_{c}\rightarrow\gamma\eta^{\prime})$ are in very nice agreement with their experiment data~\cite{Ablikim:2016uoc}. Furthermore, we find that relativistic corrections increase the individual branching ratios $\mathcal{B}(h_{c}\rightarrow\gamma\eta)$ and $\mathcal{B}(h_{c}\rightarrow\gamma\eta^{\prime})$ by about a factor of $2$, which are independent on the choice of the phenomenological parameters. This may imply that relativistic corrections are also important in other decay processes of the $P$-wave $h_{c}$. In addition, the gluonic contributions are comparable with the quark-antiquark contributions, and this is also independent on the choice of the phenomenological parameters.

\subsection{$\eta-\eta^{\prime}$ mixing}
\label{subsec:etaetapmixing}

\begin{figure}[!!htb]
\centering
      \includegraphics[width=0.5\textwidth]{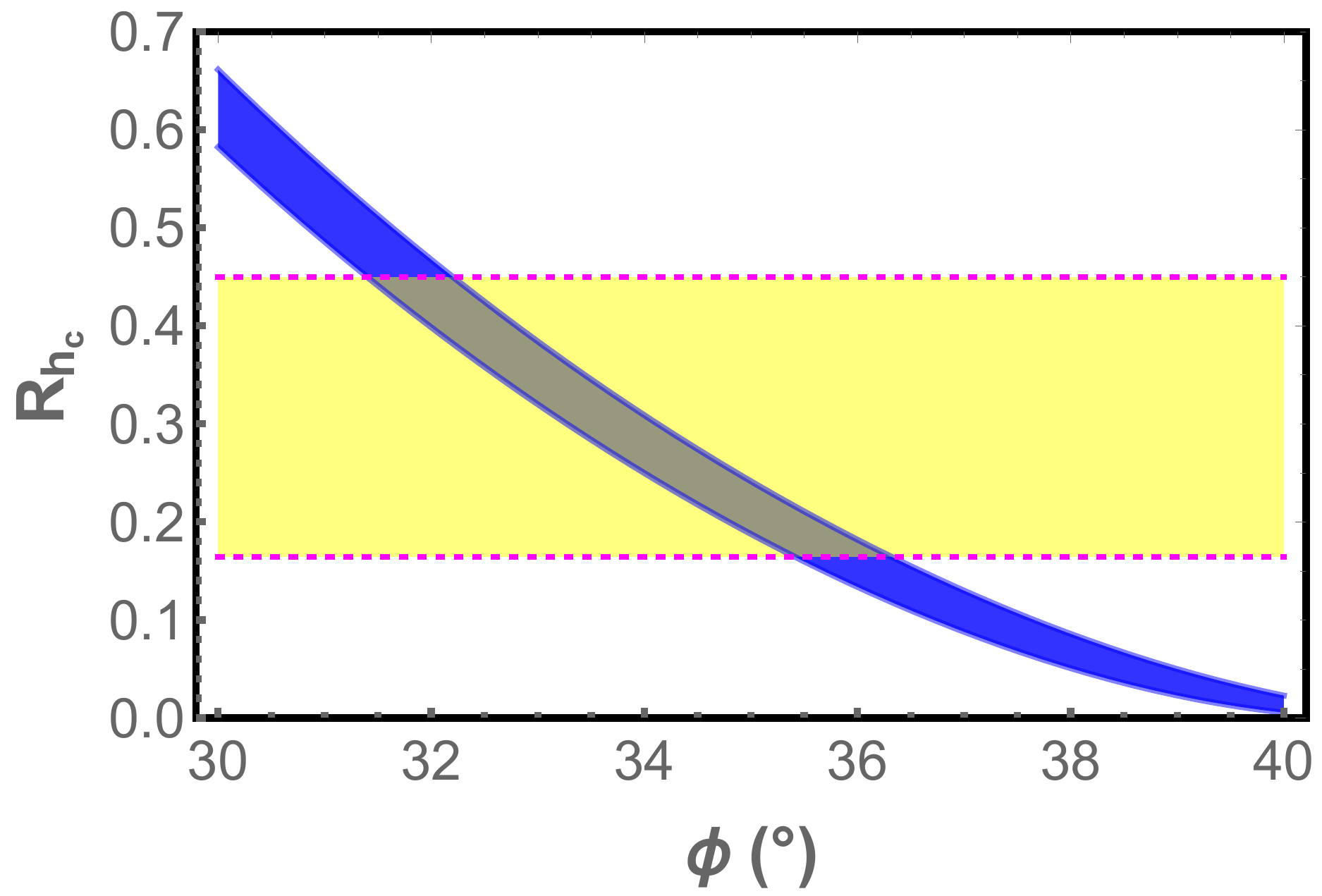}
\caption{\label{Rhc}The dependence of the ratio $R_{h_{c}}$ on the mixing angle $\phi$. The blue band is our calculated results with the uncertainties from the $\Gamma^{exp}(\eta^{(\prime)}\rightarrow\gamma\gamma)$. The yellow band denotes the experimental value of $R_{h_{c}}$ with $1 \sigma$ uncertainty.}
\end{figure}

As a cross-check, we give a prediction of the mixing angle $\phi$ in the B-S formalism. By the ratio
\begin{eqnarray}\label{rhc}
R_{h_{c}}=\frac{M^{2}-m_{\eta}^{2}}{M^{2}-m_{\eta^{\prime}}^{2}}
\frac{{\mid}H_{QCD}^{q}+H_{QCD}^{g}{\mid}^{2}_{m=m_{\eta}}}
{{\mid}H_{QCD}^{q}+H_{QCD}^{g}{\mid}^{2}_{m=m_{\eta^{\prime}}}},
\end{eqnarray}
the ratio
\begin{eqnarray}\label{rgam}
\frac{\Gamma(\eta\rightarrow\gamma\gamma)}{\Gamma(\eta^{\prime}\rightarrow\gamma\gamma)}
=\frac{m_{\eta}^{3}}{m_{\eta^{\prime}}^{3}}\left(\frac{5\sqrt{2} \frac{f_{s}}{f_{q}}-2 \tan \phi }{5\sqrt{2} \frac{f_{s}}{f_{q}} \tan \phi +2 }\right)^{2}
\end{eqnarray}
and the experimental measurements~\cite{Ablikim:2016uoc,Babusci:2012ik,Zyla:2020zbs}
\begin{eqnarray}
& &R_{h_{c}}^{exp}=(30.7\pm11.3\pm8.7)\%, \nonumber\\
& &\Gamma^{exp}(\eta^{\prime}\rightarrow\gamma\gamma)=4.34(14)~\mathrm{keV}, \nonumber\\
& &\Gamma^{exp}(\eta\rightarrow\gamma\gamma)=0.516(18)~\mathrm{keV},
\end{eqnarray}
one can obtain the mixing angle
\begin{eqnarray}
\phi&=&33.8^{\circ}\pm2.5^{\circ},
\end{eqnarray}
where the uncertainty comes mainly from the $R^{exp}_{h_{c}}$. It is worth noting that the recent lattice QCD calculations give the following values: $\phi=34^{\circ}\pm3^{\circ}$ from the UKQCD collaboration~\cite{Gregory:2011sg} and $\phi=38.8^{\circ}\pm3.3^{\circ}$ from the ETM collaboration~\cite{Ottnad:2017bjt}. Schematically, we show the dependence of the ratio $R_{h_{c}}$ on the mixing angle $\phi$ in Fig.~\ref{Rhc}. Obviously, the prediction of the mixing angle in the B-S framework is consistent with the value $\phi=33.5^{\circ}\pm0.9^{\circ}$ extracted from $\eta^{\prime}$ TFF~\cite{Escribano:2013kba}. Moreover, it is also in good agreement with our previous determinations $\phi=33.9^{\circ}\pm0.6^{\circ}$~\cite{He:2019mpy} and $\phi=33.8^{\circ}\pm2.5^{\circ}$~\cite{Fan:2019sap}, which are obtained by the ratios $R_{J/\psi}$ and $R_{h_{c}}$ without considering the relativistic corrections respectively.

It is worth noting that there are discrepancies in the determinations of the mixing angle $\phi$, and it may imply an incomplete understanding of $\eta-\eta^{\prime}$ mixing. Last but interestingly, in the same framework of perturbative QCD, we obtain a consistency check about the mixing angle by the ratio $R_{J/\psi}$ and the ratio $R_{h_{c}}$. No matter what it is, the physics associated with the $\eta-\eta^{\prime}$ mixing is very important, and certainly worth further investigations to catch more features about it.

\section{Summary}
\label{sec:conclusion}

In this work, we have revisited the $P$-wave charmonium radiative decays $h_{c}\rightarrow \gamma\eta^{(\prime)}$ in the B-S formalism, where the internal momentum of $h_{c}$ has been retained in both the soft wave function $\psi(\hat{q})$ and the hard-scattering amplitude $\mathcal{O}(\hat{q})$. The B-S wave function is employed to describe the bound-state properties of the $P$-wave charmonium $h_{c}$, while the light-cone DAs are adopted for the light mesons $\eta^{(\prime)}$.  And then the involved one-loop integrals are carried out analytically. It is found that the relativistic corrections from both the quark-antiquark content and the gluonic content of $\eta^{(\prime)}$ make significant contributions to the decay rates of $h_{c}\rightarrow\gamma\eta^{(\prime)}$. In addition, the predicted branching ratios $\mathcal{B}(h_{c}\rightarrow\gamma\eta^{(\prime)})$ are insensitive to the shapes of the $\eta^{(\prime)}$ DAs, and the gluonic contributions as well as the quark-antiquark contributions are both important in these two decay processes. What is more, by the ratio $R_{h_{c}}$, the widths $\Gamma(\eta^{(\prime)}\rightarrow\gamma\gamma)$ and their experimental values, we have obtained a consistency check about the mixing angle $\phi$ in the framework of perturbative QCD.

As aforementioned discussion, since the decay amplitude of the $\eta$ channel is proportional to the factor $(\sqrt{2}f_{q}\cos\phi-f_{s}\sin\phi)$ which could lead to large cancellations in the matrix elements, the predicted branching ratio $\mathcal{B}(h_{c}\rightarrow\gamma\eta)$ is sensitive to the angle of $\eta-\eta^{\prime}$ mixing. This means that the branching ratios are generally hard to predict precisely, but would be more efficient to determine the mixing angle in $\eta$ production and decay processes. On the other hand, by the comparison between the results without the relativistic corrections and these obtained in this work, we find that the relativistic corrections are rather significant in the exclusive $P$-wave charmonium decays $h_{c}\rightarrow\gamma\eta^{(\prime)}$. This may imply that the relativistic effects should be taken into account in the production and decay processes of the higher excited charmonia, especially for the radially-excited states with nodes contained in their wave functions. Further investigations about these issues are certainly deserved.

\section*{Acknowledgements}
This work is supported by the National Natural Science Foundation of China under Grant Nos.~11675061, 11775092 and 11435003.

\newpage

\appendix

\section*{Appendix: The expressions of the six numerators}
\label{sec:six numerators}
The expressions of the numerators $N_{i}$ ($i=1\sim6$) read
\begin{eqnarray*}
N_{1}&=&\frac{64}{1-x} \Big(M \hat{q}^{2} l^{2} \left(x^2-1\right)-2 m_{c} k\cdot\hat{q} \left(k\cdot l-p\cdot l + x K\cdot l\right)\Big)+16 M l\cdot\hat{q} l^{2} \left(x+1\right)\nonumber\\
&&+\frac{128 k\cdot\hat{q} K\cdot l^{2}}{M^3 \hat{m}_{c} \left(1-x\right)} \Big(\hat{m}_{c} \left(2 k\cdot\hat{q}+p\cdot l\right)-2 m_{c} k\cdot\hat{q}\Big)+\frac{64 K\cdot l l\cdot p \left(2 \hat{q}^{2}-l\cdot\hat{q}\right)}{M}\nonumber\\
&&-\frac{128 k\cdot\hat{q} K\cdot l l\cdot\hat{q} \left(2 m_{c} x-\hat{m}_{c} x-\hat{m}_{c}\right)}{M \hat{m}_{c} \left(1-x\right)}+\frac{64 M m_{c} l\cdot\hat{q}^{2} \left(x+1\right)}{\hat{m}_{c}}\nonumber\\
&&+\frac{64 M l\cdot\hat{q} }{\hat{m}_{c}}\Big(M m_{c} \hat{m}_{c} \left(x-1\right)+\left(x+1\right) \left(m_{c}^2 \hat{m}_{c}-2 m_{c} \hat{q}^{2}+\hat{m}_{c} \hat{q}^{2}\right)\Big)\nonumber\\
&&+\frac{128 k\cdot\hat{q} K\cdot l }{M \hat{m}_{c}}\Big(m_{c}^2 \hat{m}_{c}-2 m_{c} \hat{q}^{2}+\hat{m}_{c} \hat{q}^{2}\Big)+\frac{32 k\cdot\hat{q} K\cdot l l^{2} \left(x+3\right)}{M \left(x-1\right)},
\end{eqnarray*}
\begin{eqnarray*}
N_{2}&=&\frac{16 M x }{\hat{m}_{c}}\Big(M^2 x \hat{m}_{c} l\cdot\hat{q}+ 2 \hat{q}^{2} \left(m_{c}-\hat{m}_{c}\right) \left(2 k\cdot l+l\cdot p\right)-\hat{m}_{c} l\cdot\hat{q} \left(\left(M-2 m_{c}\right)^2-4 \hat{q}^{2}\right)\Big)\nonumber\\
&&+\frac{32 M k\cdot\hat{q}}{\hat{m}_{c} \left(x-1\right)} \Big(2 \hat{m}_{c} l\cdot p - 2 \hat{m}_{c} \left(x^{2}-2 x - 1\right) l\cdot\hat{q}+\hat{m}_{c} \left(x^{2}-3 x\right) K\cdot l - 4 m_{c} x l\cdot\hat{q}\Big)\nonumber\\
&&-\frac{32 l^{2} M }{\hat{m}_{c} \left(x-1\right)}\Big(m_{c} \left(x^2-1\right) \hat{q}^{2} + 2 x \hat{m}_{c} k\cdot\hat{q}+\hat{m}_{c} \left(x^2-1\right)\hat{q}^{2} \Big)+\frac{64 \left(x-2\right) k\cdot\hat{q} l\cdot p^{2}}{M \left(1-x\right)}\nonumber\\
&&+\frac{128 k\cdot\hat{q}^2}{M \hat{m}_{c}\left(1-x\right)} \Big(m_{c} l\cdot p +\hat{m}_{c} \left(x-2\right) l\cdot p - \hat{m}_{c}\left(1-x\right) k\cdot l\Big)+\frac{64 x k\cdot\hat{q} k\cdot l l\cdot p}{M\left(1-x\right)}\nonumber\\
&&-\frac{32 M }{\hat{m}_{c}}\Big(2 m_{c}^2 \hat{m}_{c} l\cdot\hat{q}+ m_{c} \hat{q}^{2} l\cdot p- \hat{m}_{c} \hat{q}^{2} \left(l\cdot p+2 l\cdot\hat{q}\right)\Big)-\frac{128 \left(m_{c}^2-\hat{q}^{2}\right) k\cdot\hat{q} K\cdot l}{M}\nonumber\\
&&+\frac{32 l\cdot p l\cdot\hat{q}}{M \hat{m}_{c} \left(x-1\right)} \Big(M^2 \hat{m}_{c} \left(x-1\right) x-2 M m_{c} \hat{m}_{c} \left(x-1\right)+4 \left(m_{c}-\hat{m}_{c}\right) k\cdot\hat{q}\Big)\nonumber\\
&&+\frac{64 K\cdot l l\cdot p}{M^2 \hat{m}_{c} \left(x-1\right)} \Big(M m_{c} \left(x-1\right) \hat{q}^{2}+2 m_{c} \hat{m}_{c} k\cdot\hat{q}+M \hat{m}_{c} \left(x-1\right) \hat{q}^{2}\Big)\nonumber\\
&&-\frac{128 l^{2} \left(M m_{c} \hat{m}_{c} k\cdot\hat{q}+k\cdot\hat{q}^2 \left(m_{c}-\hat{m}_{c}\right)\right)}{M \hat{m}_{c} \left(x-1\right)}+\frac{128 m_{c} k\cdot\hat{q} \left(x K\cdot l -l\cdot p\right)}{\left(x-1\right)},
\end{eqnarray*}
\begin{eqnarray*}
N_{3}&=&\frac{32 M m_{c}}{\hat{m}_{c}} \Big(2 M \hat{m}_{c} x l\cdot\hat{q} + l\cdot p \hat{q}^{2} - 2 x k\cdot l \hat{q}^{2} + x l\cdot p \hat{q}^{2} - 2 m_{c} \hat{m}_{c} \left(x+1\right) l\cdot\hat{q}\Big)\nonumber\\
&&+\frac{32 l\cdot p l\cdot\hat{q}}{M \hat{m}_{c} \left(x-1\right)} \Big(M^2 \hat{m}_{c} \left(x^{2} - x\right)-2 M m_{c} \hat{m}_{c} \left(x-1\right)+4 \left(\hat{m}_{c}-m_{c}\right) k\cdot\hat{q}\Big)\nonumber\\
&&+\frac{64 x k\cdot\hat{q} k\cdot l l\cdot p}{M \left(x-1\right)}+\frac{128 m_{c} k\cdot\hat{q} \left(x K\cdot l-l\cdot p\right)}{\left(x-1\right)}-\frac{128 \left(m_{c}^2-\hat{q}^{2}\right) k\cdot\hat{q} K\cdot l}{M}\nonumber\\
&&+16 M \Big(M^2 \left(x^{2}-x\right) l\cdot\hat{q}+2 \hat{q}^{2} \left(2 x k\cdot l+2 x l\cdot\hat{q}+2 l\cdot\hat{q}+ \left(x-1\right) l\cdot p\right)\Big)\nonumber\\
&&+\frac{32 M  k\cdot\hat{q}}{\left(x-1\right)}\Big(\left(x^{2}-3 x\right) k\cdot l+\left(x^{2}-3 x+2\right) l\cdot p+2 \left(x^{2}-2 x -1\right) l\cdot\hat{q}\Big)\nonumber\\
&&+\frac{128 l^{2} k\cdot\hat{q} }{M \hat{m}_{c} \left(x-1\right)}\Big(M m_{c} \hat{m}_{c} - \left(m_{c}-\hat{m}_{c}\right)k\cdot\hat{q}\Big)+\frac{64 \left(x-2\right) k\cdot\hat{q} l\cdot p^{2}}{M \left(x-1\right)}\nonumber\\
&&+\frac{64 K\cdot l l\cdot p}{M^2 \hat{m}_{c} \left(x-1\right)} \Big(M m_{c} \left(x-1\right) \hat{q}^{2}-2 m_{c} \hat{m}_{c} k\cdot\hat{q}+M \hat{m}_{c} \left(x-1\right) \hat{q}^{2}\Big)\nonumber\\
&&-\frac{32 l^{2} M }{\hat{m}_{c} \left(x-1\right)}\Big(m_{c} \left(x^2-1\right) \hat{q}^{2}+\hat{m}_{c} \left(x^2-1\right) \hat{q}^{2}-2 \hat{m}_{c} x k\cdot\hat{q}\Big)\nonumber\\
&&+\frac{128 k\cdot\hat{q}^2 }{M \hat{m}_{c} \left(x-1\right)}\Big(m_{c} l\cdot p-\hat{m}_{c} \left(k\cdot l+2 l\cdot p\right) + x \hat{m}_{c} K\cdot l\Big)\nonumber\\
&&+\frac{128 M m_{c} x k\cdot\hat{q} l\cdot\hat{q}}{\hat{m}_{c} \left(x-1\right)},
\end{eqnarray*}
\begin{eqnarray*}
N_{4}&=&\frac{8 i \left(M^2 (x \xi+x-1)-2 M m_{c} (\xi-1)-4 m_{c}^2\right)}{M (x-1)}k\cdot\hat{q}+\frac{32 i}{M (x-1)}k\cdot\hat{q}\, \hat{q}^{2}\nonumber\\
& &+\frac{16 i (m_{c} (x+1) (\xi-1)-\hat{m}_{c} (x \xi+x+\xi-3))}{M \hat{m}_{c} (x-1)^2}(k\cdot\hat{q})^{2}\nonumber\\
&&+\frac{4 i M (x-1) (m_{c} (\xi-1)+\hat{m}_{c} (\xi+1))}{\hat{m}_{c}}\hat{q}^{2},
\end{eqnarray*}
\begin{eqnarray*}
N_{5}&=&-\frac{8 i  \left(M^2 (x (\xi-1)+1)-2 M m_{c} (\xi+1)+4 m_{c}^2\right)}{M (x-1)}k\cdot\hat{q}+\frac{32 i}{M (x-1)}k\cdot\hat{q}\, \hat{q}^{2}\nonumber\\
& &+\frac{16 i  (m_{c} (x+1) (\xi+1)-\hat{m}_{c} (x (\xi-1)+\xi+3))}{M \hat{m}_{c} (x-1)^2}(k\cdot\hat{q})^{2}\nonumber\\
&&+\frac{4 i M  (x-1) (m_{c} (\xi+1)+\hat{m}_{c} (\xi-1))}{\hat{m}_{c}}\hat{q}^{2},
\end{eqnarray*}
\begin{eqnarray*}
N_{6}&=&-\frac{8 i \left(M^2 \xi^2-4 m_{c}^2\right)}{M (x-1)}k\cdot\hat{q}-\frac{32 i (2 m_{c}-\hat{m}_{c})}{M \hat{m}_{c} (x-1)}k\cdot\hat{q}\, \hat{q}^{2}\nonumber\\
&&-\frac{32 i \xi (2 m_{c}-\hat{m}_{c} (x+1))}{M \hat{m}_{c} (x-1)^2}(k\cdot\hat{q})^{2}+8 i M (x-1) \xi\hat{q}^{2}
\end{eqnarray*}
with $x=m^{2}/M^{2}$.

\newpage


\bibliographystyle{JHEP}
\bibliography{hejk}

\end{document}